# The Chiral Anomaly and Ultrahigh Mobility in Crystalline HfTe$_5$

Huichao Wang,[1,2] Chao-Kai Li,[1,2] Haiwen Liu,[1,2] Jiaqiang Yan,[3,4] Junfeng Wang,[5] Jun Liu,[6] Ziquan Lin,[5] Yanan Li,[1,2] Yong Wang,[6] Liang Li,[5] David Mandrus,[3,4] X. C. Xie,[1,2] Ji Feng,[1,2] and Jian Wang[1,2,*]

[1]*International Center for Quantum Materials, School of Physics, Peking University, Beijing 100871, China*
[2]*Collaborative Innovation Center of Quantum Matter, Beijing 100871, China*
[3]*Department of Materials Science and Engineering, University of Tennessee, Knoxville, Tennessee 37996, USA*
[4]*Materials Science and Technology Division, Oak Ridge National Laboratory, Oak Ridge, Tennessee 37831, USA*
[5]*Wuhan National High Magnetic Field Center, Huazhong University of Science and Technology, Wuhan 430074, China*
[6]*Center of Electron Microscopy, State Key Laboratory of Silicon Materials, Department of Materials Science and Engineering, Zhejiang University, Hangzhou, 310027, China*

HfTe$_5$ is predicted to be a promising platform for studying topological phases. Here through an electrical transport study, we present the first observation of chiral anomaly and ultrahigh mobility in HfTe$_5$ crystals. Negative magneto-resistivity in HfTe$_5$ is observed when the external magnetic and electrical fields are parallel (**B**//**E**) and quickly disappears once **B** deviates from the direction of **E**. Quantitative fitting further confirms the chiral anomaly as the underlying physics. Moreover, by analyzing the conductivity tensors of longitudinal and Hall traces, ultrahigh mobility and ultralow carrier density are revealed in HfTe$_5$, which paves the way for potential electronic applications.

Chiral anomaly is a quantum anomaly phenomenon that breaks the chiral symmetry and leads to the non-conservation of chiral current [1,2]. This anomaly was proposed to be observed in lattice system in 1983 [3]. Recently, the study of Weyl fermions pushes forward the realization of chiral anomaly in crystals [4-6]. In Dirac/Weyl semimetals, the axial current is non-conserved due to the chiral anomaly, and further leads to charge pumping effect between the Weyl nodes with opposite chirality. This anomaly effect is suggested to give rise to negative magneto-resistivity when the magnetic and electrical fields are parallel (**B**//**E**) [4]. Related experimental evidences have been intensively pursued in various condensed matter systems [7-20].

HfTe$_5$ is a layered material with a van der Waals coupling between the individual layers. It crystallizes in an orthorhombic structure with the space group *Cmcm* [21]. The trigonal prismatic chains of "HfTe$_3$" along the *a*-axis are linked along the *c*-axis via parallel zigzag chains of "Te$_2$", forming a two-dimensional (2D) sheet of HfTe$_5$ in the *a-c* plane. The sheets of HfTe$_5$ stack along the *b*-axis. HfTe$_5$ has been previously studied for its resistivity anomaly [22,23], thermoelectric properties [24] and quantum oscillations [25,26]. Previous *ab initio* study has predicted that HfTe$_5$ and ZrTe$_5$ crystals are located near the phase boundary between weak and strong topological insulators [27]. Successive ARPES results reveal that ZrTe$_5$ is a three-dimensional Dirac semimetal [13]. Moreover, recent experimental investigations have shown that ZrTe$_5$ possesses many peculiar features, such as the B$^{1/2}$-dependence of inter-Landau-level resonance [28,29] and non-trivial Berry phase [29], which confirm the existence of Dirac fermions in the system. Particularly，the chiral anomaly is observed in magneto-transport measurements of ZrTe$_5$ [13]. However, theoretically predicted topological material HfTe$_5$ has rarely been confirmed and studied experimentally.

Here we present the first electrical transport evidence for the chiral anomaly and the ultrahigh mobility in HfTe$_5$ crystals.

In specific, anomalous negative magneto-resistivity is observed when the magnetic field **B** is aligned along the direction of electrical field **E**. It is quite sensitive to the orientation of **B** relative to **E** and disappears rapidly when **B** deviates from **E**. Quantitative analyses suggest the negative magneto-resistivity originates from the chiral anomaly. Moreover, Hall trace of $HfTe_5$ crystal is studied. By analyzing both the longitudinal and Hall conductivities in a two-carrier model, we find $HfTe_5$ has ultrahigh mobility and ultralow carrier density especially at low temperatures. The Hall trace also shows that in our $HfTe_5$ samples only hole type carriers contribute to the Fermi surface, which is consistent with the previously calculated band structure of $HfTe_5$ [27].

$HfTe_5$ crystals were grown out of Te-flux using a Canfield crucible set (CCS) [30]. The self-flux growth yields high quality crystals as in the case of $WTe_2$ [31]. Hf pieces and Te shots in an atomic ratio of 1:99 were loaded into the CCS and then sealed in a silica ampoule under vacuum. The sealed ampoule was heated to 800 °C and kept for 8 hours to homogenize the melt, furnace cooled to 600 °C, and then cooled down to 460 °C in 48 hours. $HfTe_5$ crystals were isolated from Te flux by centrifuging once the furnace temperature reached 460 °C. Typical $HfTe_5$ samples are about 10-20 mm long with the other two dimensions in the range of 0.01 mm-0.6 mm. The longest direction is always along the crystallographic *a*-axis.

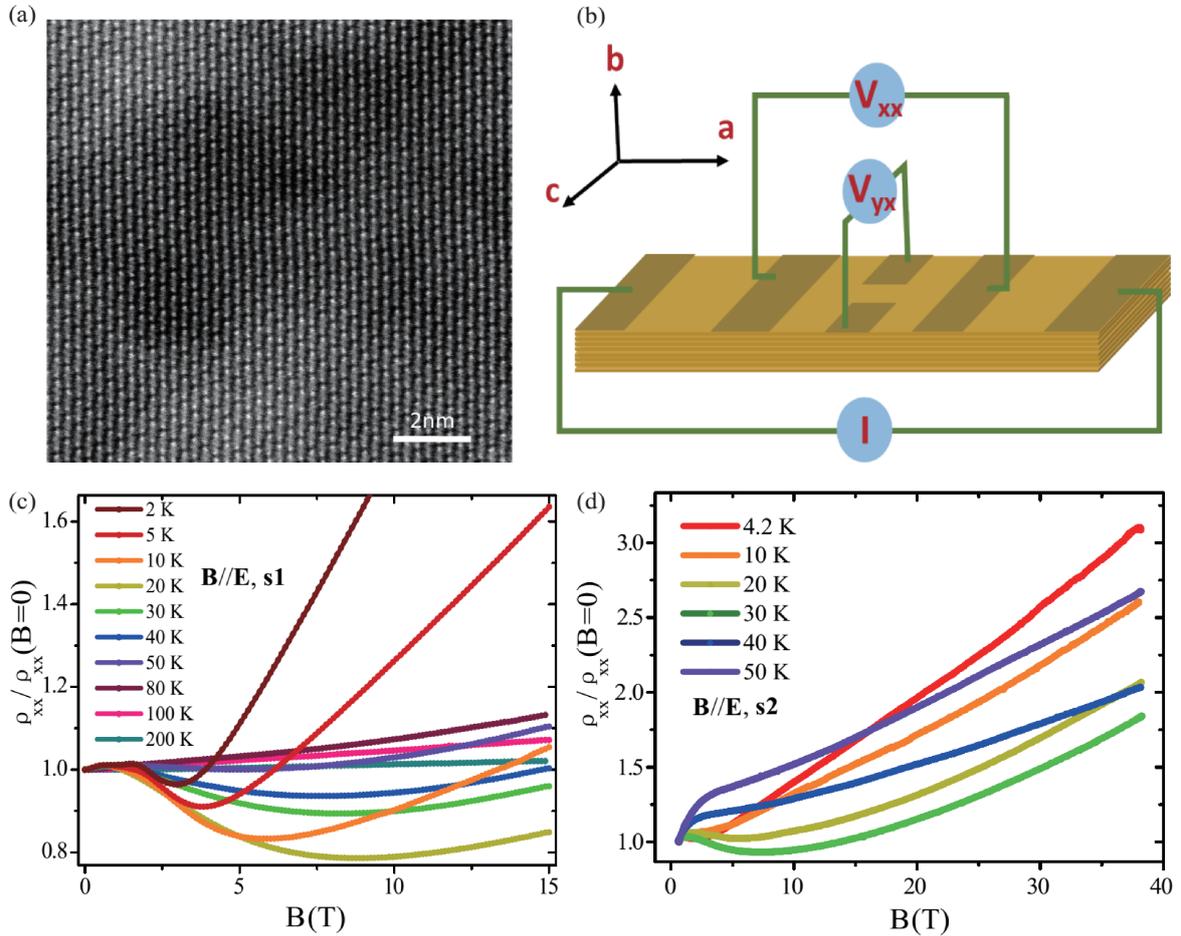

FIG. 1 (a) HAADF STEM image of a typical $HfTe_5$ crystal. Scale bar represents 2 nm. (b) A schematic structure for the electrical transport measurements of $HfTe_5$. Standard four(six)-probe method is used for the resistivity (resistivity and Hall trace) measurement. (c) Temperature dependence of the normalized magneto-resistivity in $HfTe_5$ sample 1 (s1) when **B**//**E**. (d) Normalized magneto-resistivity of sample 2 (s2) extended to 38 T at selected temperatures with **B**//**E**.

We use a FEI TITAN Cs-corrected cross-sectional STEM operating at 200 kV to examine the HfTe$_5$ crystal. As shown in Fig. 1(a), the atomic layer-by-layer high angle annular dark field (HAADF) STEM image manifests a high-quality nature of the HfTe$_5$ crystal. Electrical transport measurements are mainly conducted in a 16T-Physics Property Measurement System (PPMS-16T) from Quantum Design. High magnetic field study extended to 38 T is performed in the Wuhan National High Magnetic Field Center. Standard four(six)-probe method is used for measuring resistivity (resistivity and Hall trace). Electrical contacts were made by using 25-μm-diameter Au wires bonded to the crystal with silver paint. A schema of the measurement configuration is shown in Fig. 1(b). The external electrical field **E** is always along the *a*-axis in our work. Sample is rotated to form different orientation relative to the magnetic field. Tens of samples from the same batch are studied and typical results are shown in the main text.

We explore the chiral anomaly of HfTe$_5$ in the measurement configuration when both magnetic and electrical fields are aligned along the *a*-axis. Figure 1(c) shows the normalized magneto-conductivity of sample 1 (s1) at selected temperatures when **B**//**E**. In the very weak **B** around **B**=0, the sample shows positive magneto-resistivity signaling weak anti-localization effect induced by the strong spin-orbit coupling. Surprisingly, an anomalous negative magneto-resistivity is observed at low temperatures, which almost disappears at 50 K. At the large **B** regime, the magneto-resistivity becomes positive which may be due to the electron-electron interaction or the misalignment of excitation current with external magnetic field. With similar results to s1 below 15 T, s2 is further measured in pulsed magnetic field up to 38 T with **B**//**E**, as shown in Fig. 1(d). At the high magnetic field above 15 T, s2 shows positive magneto-resistivity without any oscillation. Specially, linear magneto-resistivity is observed at 4.2 K in large magnetic field regime, similar to the observation in Dirac semimetal Cd$_3$As$_2$ system [9,32,33].

For chiral anomaly induced negative magneto-resistivity, it is predicted to be most evident when **B**//**E**. To test the property of our measured negative magneto-resistivity, we rotate s1 in the *a-b* plane at 2 K. When the *b*-axis of s1 is aligned to the magnetic field direction (θ=0º), i.e. **B** is perpendicular to **E**, the magneto-resistivity is always positive and saturated at high magnetic fields (Fig. 2(a)). With sample tilted to a larger θ, the magneto-resistivity becomes smaller and decreases abruptly around θ=90º. More detailed measurement results around θ=90º are shown in Fig. 2(b). The negative magneto-resistivity is most evident when **B**//**E** and can survive only if the relative orientation of **B** and **E** is smaller than 1°. Magneto-resistivity of s1 below 10 T as a function of the perpendicular field component, Bcosθ, is shown in Fig. 2(c). We find that the results for small θ (<65°) can be perfectly collapsed onto one single curve, suggesting a quasi-two-dimensional feature. However, the ρ$_{xx}$ (**B**) curves obtained at large θ (65°<θ<90°) deviate from the scaling. When θ is close to 90° (**B**//**E)**, the deviation becomes quite apparent, indicating other certain underlying effect.

The study in high magnetic field up to 38 T shows that the negative magneto-resistivity is not a part of Shubnikov–de Haas (SdH) oscillation. In addition, the corresponding magnetic field for resistivity minimum is shifting at different temperatures, which is also inconsistent with the SdH behavior. On the other hand, as the angle-dependent results show in Fig. 2(b), the negative magneto-resistivity can only be observed when **B** is aligned to within 1° of the direction of **E**. Thus, the weak localization effect which is not so sensitive to the direction of **B** with respect to **E** can be excluded. All these peculiar features in our observation point to a highly possible mechanism, the chiral anomaly. For further confirmation, quantitative analyses are carried out by fitting the experimental results in Fig. 2(d) with a simplified semi-classical formula [7]

$$\sigma_{xx} = (\sigma_0 - a\sqrt{B})(1 + C_w B^2), \tag{1}$$

where the first term is related to a weak anti-localization correction with coefficient a > 0 and the second term is associated with the chiral anomaly. σ$_0$ is the zero field conductivity and C$_w$ is a positive parameter. Black lines in Fig. 2(d) are produced by Equation (1). The nice match between theoretical fitting and experimental results from 0 to 3 T provides a strong evidence for the chiral anomaly mechanism. The estimated $C_w$ from fitting (see Supplemental Material) is monotonically decreasing from 0.014±0.001 at 2 K to 0.002±(<0.001) at 50 K, revealing a reasonable temperature dependence of the chiral anomaly.

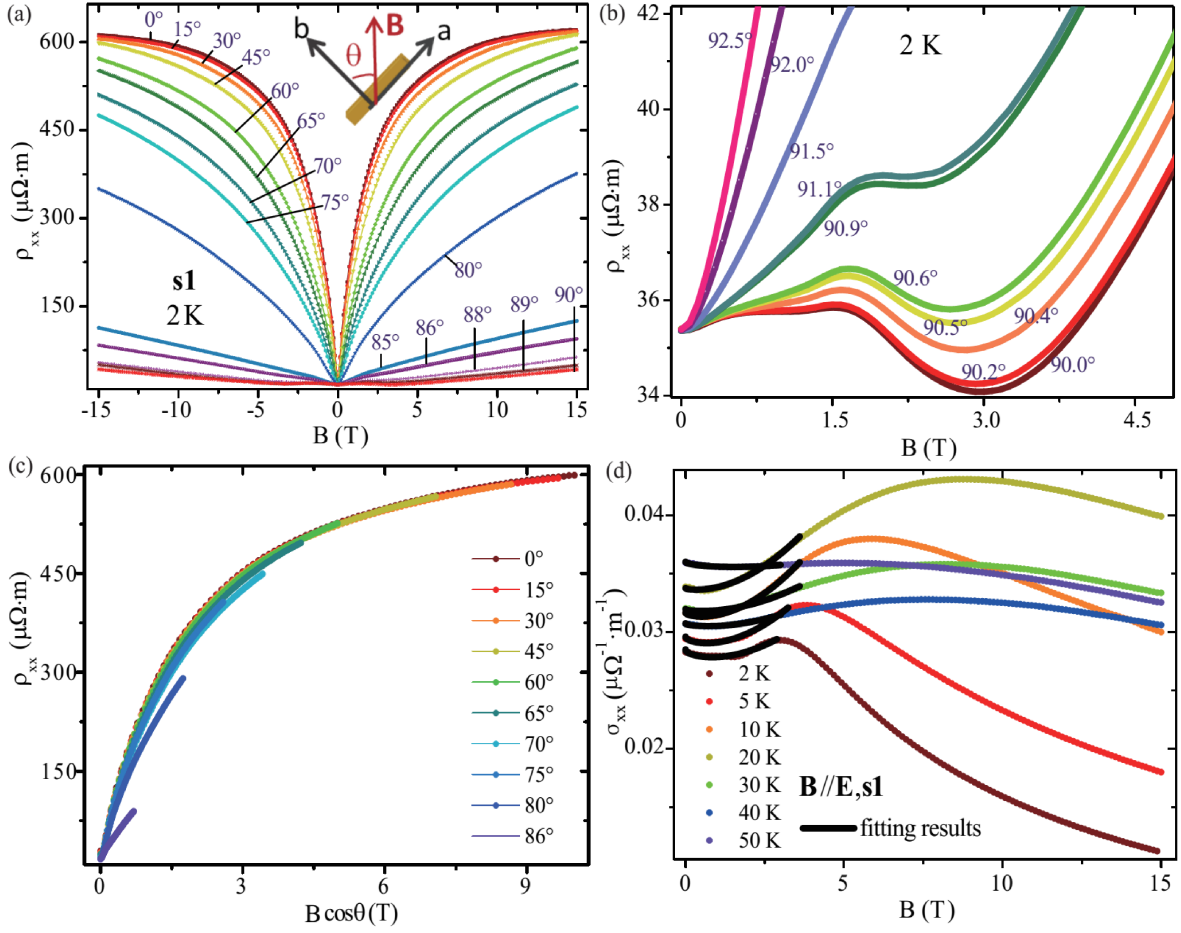

FIG. 2 (a) Magneto-resistivity $\rho_{xx}$ vs. **B** at selected angles θ when s1 is rotated in the *a-b* plane. The magnetic field is along *b*-axis (**B**⊥**E**) for θ=0° and along *a*-axis (**B**//**E**) for θ=90°. (b) Anomalous negative magneto-resistivity in s1 when the external magnetic and electrical fields are nearly parallel. (c) Magneto-resistivity vs. the perpendicular component of magnetic field **B**cosθ. Results at small θ indicate a quasi-two-dimensional nature. (d) Quantitative fitting for the magneto-conductivity of s1 at different temperatures from 0 to 3 T when **B**//**E**. Dots are experimental data and black lines are fitting results with Equation (1).

To further understand the properties of HfTe$_5$, Hall effect is studied in sample 3 (s3) as shown in Fig. 3. Fig. 3(a) shows the magneto-resistivity $\rho_{xx}$ at selected temperatures under an external perpendicular magnetic field (**B**//*b* axis). The positive magneto-resistivity at 2 K is as large as 6000 % in an applied filed of 15 T. Even at room temperature 300 K, the value of ρ (15 T) /ρ (0 T) is about 150 %. Hall resistivity $\rho_{yx}$ ($V_{yx}/I$) of HfTe$_5$ crystal is shown in Fig. 3(b) and positive Hall coefficient is observed. At the base temperature 2 K, $\rho_{yx}$ first increases and then decreases with the increasing magnetic field. The decreasing behavior is weakened by the increased temperatures. At a temperature above 80 K, we observe a monotonically increased $\rho_{yx}$ curve with changing slope in magnetic field up to 15 T.

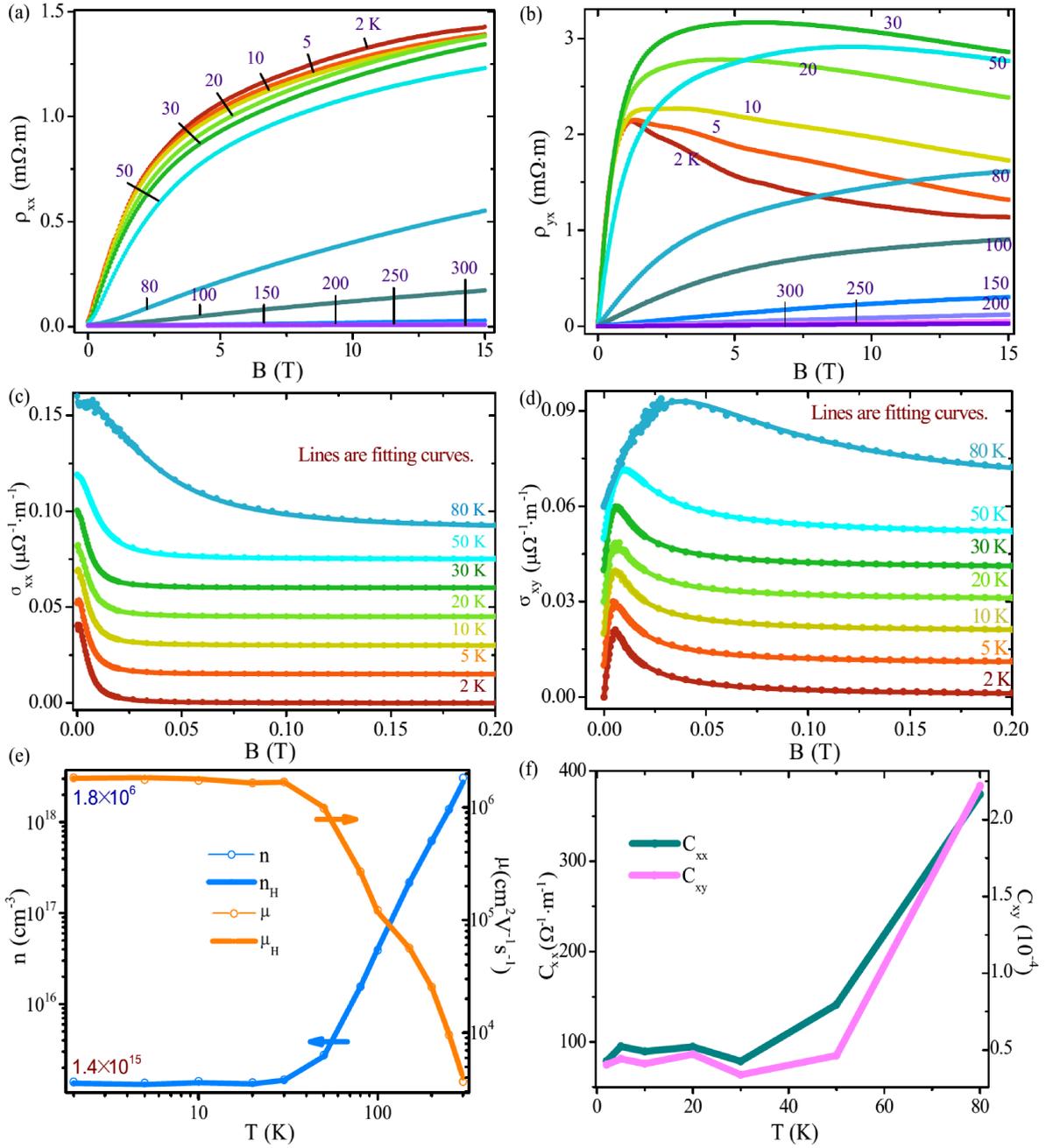

FIG. 3 Measured longitudinal resistivity (a) and Hall resistivity (b) of sample 3 (s3). (c) and (d) are corresponding conductivity tensors. Dots are experimental results and lines are fitting curves in a two-carrier model. (e) Temperature dependence of the carrier density and mobility of the dominated carriers, estimated from analyses for the conductivity tensors independently. (f) Fitting parameters $C_{xx}$ and $C_{xy}$ at low temperatures.

With the measured $\rho_{xx}$ and $\rho_{yx}$, we show the longitudinal conductivities $\sigma_{xx}$ and Hall conductivities $\sigma_{xy}$ of s3 in Figs. 3(c) and 3(d), respectively. The curves at temperatures above 2 K are shifted for clarity. In a two-carrier model, $\sigma_{xx}$ and $\sigma_{xy}$ data are fitted with the following formula [34]

$$\sigma_{xx} = \frac{ne\mu}{1+(\mu B)^2} + C_{xx},  \quad (2)$$

$$\sigma_{xy} = n_H e \mu_H^2 B \left[ \frac{1}{1+(\mu_H B)^2} + C_{xy} \right]. \quad (3)$$

Here n ($n_H$) and μ ($\mu_H$) is the density and mobility for the high-mobility carriers deduced from $\sigma_{xx}$ ($\sigma_{xy}$), respectively. $C_{xx}$ ($C_{xy}$) denotes the low-mobility components estimated from $\sigma_{xx}$ ($\sigma_{xy}$). In Figs. 3(c) and 3(d), fitting lines show excellent agreement with the experimental dots. Fitting parameters (see Supplemental Material) as a function of temperature are shown in Figs. 3(e) and 3(f). We can see from Fig. 3(e) the carrier density n and $n_H$ agrees very well with each other, though they are obtained from the analyses of $\sigma_{xx}$ and $\sigma_{xy}$ independently. The mobility μ and $\mu_H$ also coincide perfectly, which verifies the validity of our analyses within the two-carrier model. $C_{xx}$ and $C_{xy}$ related to the low-mobility carriers are displayed in Fig. 3(f). We analyze the low mobility carrier and deduce the respective $\mu_L$ from $C_{xx}$ and $C_{xy}$, which satisfies $(\mu_L \mathbf{B})^2 \ll 1$ in our fitting regime from 0 to 0.2 T for temperatures below 80 K, justifying the applicability of Equations (2) and (3).

As shown in Fig. 3(e), the carrier density of HfTe$_5$ at 2 K is estimated to be n=1.40±0.006×10$^{15}$ /cm$^3$ ($n_H$=1.35±0.006×10$^{15}$ /cm$^3$). This maps a very small Fermi surface, as estimated in previous study [24,25]. As the temperature is increased, the carrier density grows slightly and shows a clear increase above 100 K. Besides, the mobility of the dominant carriers in HfTe$_5$ is decreasing with increasing temperature. At 2 K, the mobility is remarkably high as μ=1.84±0.01×10$^6$ cm$^2$/V/s ($\mu_H$=1.82±0.01×10$^6$ cm$^2$/V/s), comparable to that of Dirac/Weyl fermions in Cd$_3$As$_2$ [9,32] and TaAs family [10-15]. It is worth noting that this is about two orders of magnitude higher than that of ZrTe$_5$ [29]. Even at room temperature, the mobility of our HfTe$_5$ crystal is as high as 3700 cm$^2$/V/s, which is significant for the potential electronic applications.

As described above, ZrTe$_5$ has been examined by ARPES result, magneto-optical spectroscopy and magneto-transport measurements [13,28,29]. The nontrivial topological properties of ZrTe$_5$ have been revealed. We know that the crystalline structure and energy band structure of HfTe$_5$ resemble those of ZrTe$_5$. From the Hall traces and two-carrier model analyses, we find that hole type carriers dominate the electrical transport in our HfTe$_5$ samples. The single type hole carrier is consistent with the previously calculated band structures in Ref. [27], in which the carrier type is either electron or hole for a given Fermi level. Our transport measurements have demonstrated that the distinct chiral anomaly originates from the topological property of HfTe$_5$, which is similar to that in Na$_3$Bi and Cd$_3$As$_2$ systems [8,17,18].

The experimental results are fully consistent with the 3D Dirac semimetal scenario, with finite Fermi surface. It may be remarked that observed quantum chiral anomaly in the transport measurements with a finite Fermi surface does not shed light on the existence of quasiparticle gap at very low energy. We therefore examine the question of gaplessness of truly low-energy excitations, which is not protected by symmetry anyway, using band structure calculation. The first-principles calculation is performed with the VASP package [35]. The projector-augmented wave pseudopotentials are employed with Perdew-Burke-Ernzerhof exchange-correlation functional. The plane-wave energy cutoff is 330 eV and the Brillouin zone is sampled with an 11×11×7 grid. The lattice is relaxed with a force threshold of 0.01 eV/Å and with the optB86b functional [36,37] in order to take into account the Van der Waals interaction.

Our first-principles calculation shows that HfTe$_5$ is a small gap topological insulator. The results are summarized in Fig. 4. The band structure without spin-orbit coupling (SOC) indicates that HfTe$_5$ is a semimetal. However, when the SOC is included, the low-energy spectrum becomes fully gapped with an indirect gap of 18.4 meV, as indicated both from the band structure and density of states. The computed band gap is comparable to the exponential fitting of the ρ-T curve in the region of intermediate temperature, as shown in the insets of Fig. 4(b). The resistance peak may come from the competition of its surface metallic states and thermally induced carriers in the insulating bulk. We have also calculated the Z$_2$ topological invariant, and the result shows that it is a 3D strong topological insulator given the experimental structure. The coexistence of bulk states and surface states may also lend support to the validity of the two-carrier model analysis of the Hall measurement. The consistency between experimental results and computed band structure with a gap therefore leads us to discuss the interesting possibility of chiral anomaly in a small gap topological insulator. As justified in Ref. [6], the chiral symmetry is always approximate in lattice systems due to the mass term, and the magnetic field induced chiral anomaly exists only when the chiral charge relaxation time $\tau_a$ greatly exceeds the mean free time $\tau_0$. Thus, in topological insulators with small energy gap and high Fermi energy, the characteristic time $\tau_a$ can also be much larger than $\tau_0$, which leads to the chiral anomaly observed in HfTe$_5$ samples. This argument is valid, with or without a band gap, and in the event the system is driven to a weak topological insulator via further band inversion by, for example, strain or lattice relaxation.

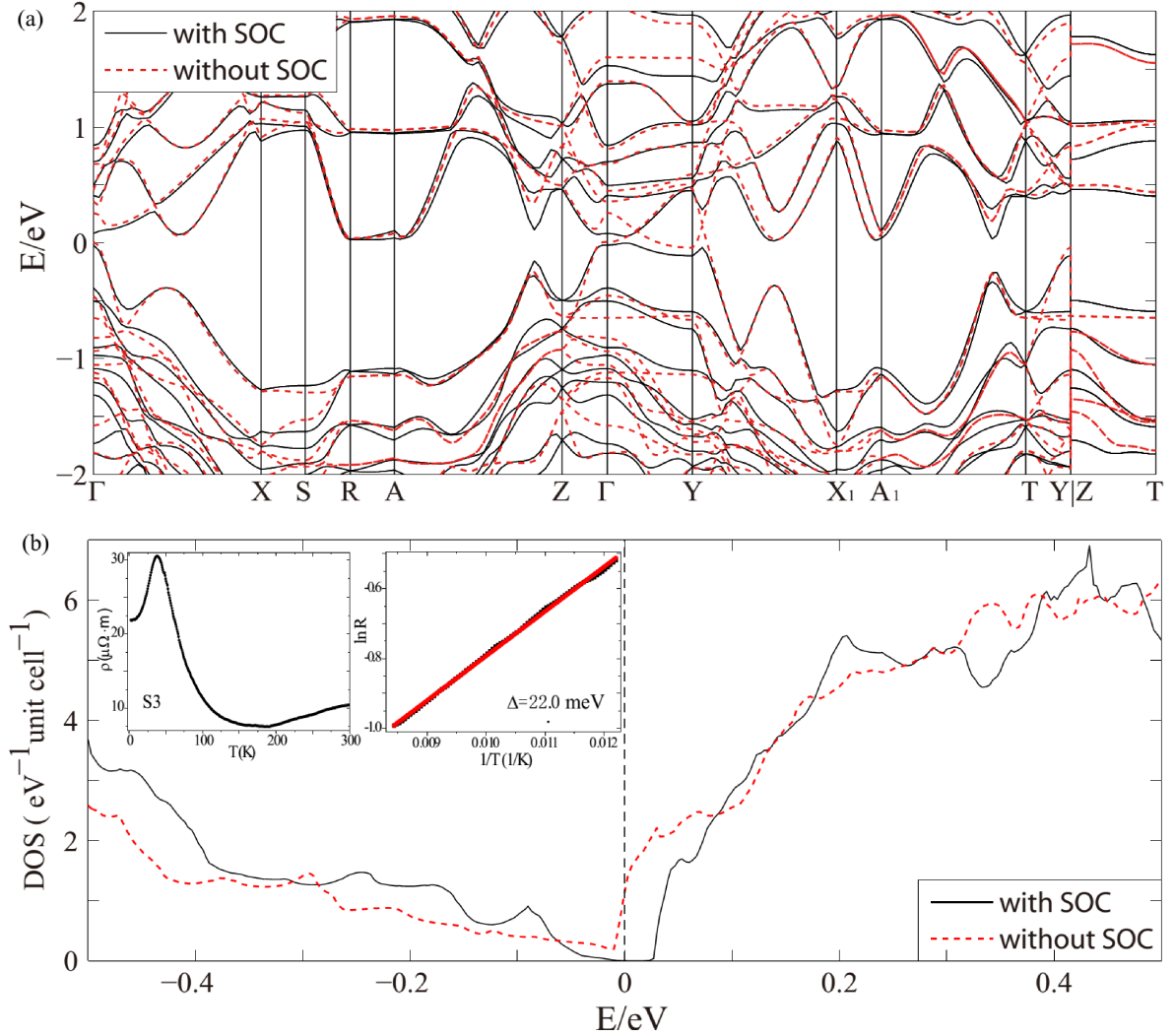

FIG. 4 (a) The calculated band structure of HfTe$_5$ with (black solid curves) and without (red dotted curves) spin-orbit coupling (SOC). The Brillouin zone path is chosen as the same with [38]. The Fermi energy is shifted to 0 eV. (b) The calculated density of states (DOS) of HfTe$_5$ with (black solid curve) and without (red dotted curve) SOC. Both the band structure and DOS results indicate an SOC-induced semimetal-insulator transition. Left inset shows the resistance vs. temperature curve of s3. The peak anomaly may be due to the competition between the metallic surface states and thermal excitations from bulk states. Right inset shows a ln R vs. 1/T plot in the temperature range from 82 K to 125 K for the same sample. This allows us to extract a gap ~ 22.0 meV through ln R ~ $\Delta/2k_BT$, close to the calculated value of 18.4 meV.

In summary, by performing systematic electrical transport measurements, we first observed the evidence for chiral anomaly and ultrahigh mobility in HfTe$_5$ crystals. Particularly, an anomalous negative magneto-resistivity is observable when the magnetic field **B** and electrical field **E** are aligned parallel. It is most evident when **B**//**E** and sensitive to the orientation of **B** relative to **E**. Our quantitative analyses identify the chiral anomaly as the underlying origin of negative magneto-resistivity. Furthermore, we find HfTe$_5$ has ultrahigh mobility and ultralow carrier density by analyzing both the longitudinal and Hall conductivities in a two-carrier model, indicating potential application in electronics.

We acknowledge Yongjie Liu for the help in the pulsed magnetic field measurements and we thank Honglie Ning for valuable discussions. This work was financially supported by the National Basic Research Program of China (Grant Nos 2013CB934600, 2012CB921300, and 2013CB921900), and the Open Project Program of the Pulsed High Magnetic Field Facility (Grant No. PHMFF2015002), Huazhong University of Science and Technology. D.G.M and J.-Q.Y acknowledge support from NSF DMR 1410428.

Huichao Wang and Chao-Kai Li contributed equally to this work.

*jianwangphysics@pku.edu.cn

# Supplemental Material for

# The Chiral Anomaly and Ultrahigh Mobility in Crystalline HfTe$_5$

Huichao Wang,[1,2] Chao-Kai Li,[1,2] Haiwen Liu,[1,2] Jiaqiang Yan,[3,4] Junfeng Wang,[5] Jun Liu,[6] Ziquan Lin,[5] Yanan Li,[1,2] Yong Wang,[6] Liang Li,[5] David Mandrus,[3,4] X. C. Xie,[1,2] Ji Feng,[1,2] and Jian Wang[1,2,*]

[1]*International Center for Quantum Materials, School of Physics, Peking University, Beijing 100871, China*
[2]*Collaborative Innovation Center of Quantum Matter, Beijing 100871, China*
[3]*Department of Materials Science and Engineering, University of Tennessee, Knoxville, Tennessee 37996, USA*
[4]*Materials Science and Technology Division, Oak Ridge National Laboratory, Oak Ridge, Tennessee 37831, USA*
[5]*Wuhan National High Magnetic Field Center, Huazhong University of Science and Technology, Wuhan 430074, China*
[6]*Center of Electron Microscopy, State Key Laboratory of Silicon Materials, Department of Materials Science and Engineering, Zhejiang University, Hangzhou, 310027, China*

Huichao Wang and Chao-Kai Li contributed equally to this work.
*jianwangphysics@pku.edu.cn

| T | $\sigma_0$ | a | $C_w$ |
|---|---|---|---|
| K | $\Omega^{-1}\cdot m^{-1}$ | $\Omega^{-1}\cdot m^{-1}\cdot T^{-1/2}$ | $T^{-2}$ |
| 2 | 28856±127 | 1396±173 | 0.014±0.001 |
| 5 | 29618±65 | 888±82 | 0.014±(<0.001) |
| 10 | 31689±44 | 643±53 | 0.014±(<0.001) |
| 20 | 33783±80 | 311±95 | 0.012±(<0.001) |
| 30 | 31975±53 | 292±62 | 0.006±(<0.001) |
| 40 | 30759±23 | 374±30 | 0.005±(<0.001) |
| 50 | 36039±23 | 474±30 | 0.002±(<0.001) |

TABLE 1: The fitting parameters for the longitudinal conductivity of HfTe$_5$ as shown in Fig. 2(d) in the main text (sample 1).

| T (K) | n ($\times 10^{21} m^{-3}$) | μ ($m^2/V/s$) | $C_{xx}$ ($\Omega^{-1}\cdot m^{-1}$) | $n_H$ ($\times 10^{21} m^{-3}$) | $\mu_H$ ($m^2/V/s$) | $C_{xy}$ |
|---|---|---|---|---|---|---|
| 2 | 1.40±(<0.01) | 183.5±1.0 | 78.8±27.1 | 1.35±(<0.01) | 181.5±1.0 | 4.1±0.5E-5 |
| 5 | 1.36±(<0.01) | 177.1±0.8 | 95.0±20.3 | 1.31±(<0.01) | 184.0±0.7 | 4.4±0.4E-5 |
| 10 | 1.42±(<0.01) | 173.3±0.5 | 89.5±13.9 | 1.37±(<0.01) | 178.9±0.6 | 4.1±0.3E-5 |
| 20 | 1.37±(<0.01) | 165.3±0.6 | 94.7±17.6 | 1.33±(<0.01) | 165.3±1.0 | 4.7±0.6E-5 |
| 30 | 1.47±(<0.01) | 168.2±0.3 | 78.7±9.0 | 1.46±(<0.01) | 168.0±0.3 | 3.4±0.2E-5 |
| 50 | 2.75±(<0.01) | 98.3±0.2 | 141.3±11.2 | 2.68±(<0.01) | 100.2±0.1 | 4.7±0.2E-5 |
| 80 | 15.50±0.08 | 26.9±0.2 | 374.3±67.6 | 15.48±0.10 | 26.6±0.1 | 2.2±0.5E-4 |
| 100 | 39.24±0.21 | 12.2±(<0.1) | 672.0±75.4 | 40.47±0.13 | 11.8±(<0.1) | 6.8±0.8E-4 |
| 150 | 216.53±0.15 | 5.7±(<0.1) | 929.8±25.3 | 217.34±0.16 | 5.7±(<0.1) | 3.7±0.1E-4 |
| 200 | 621.93±0.49 | 2.6±(<0.1) | 2326.3±51.3 | 621.28±0.38 | 2.6±(<0.1) | 9.8±0.3E-4 |
| 250 | 1377.03±1.63 | 1.0±(<0.1) | 4377.6±97.2 | 1361.76±2.89 | 1.0±(<0.1) | 31.3±3.1E-4 |
| 300 | 2887.48±2.91 | 0.4±(<0.1) | 4752.0±47.4 | 2749.04±43.12 | 0.4±(<0.1) | 112.9±48.5E-4 |

TABLE 2: The fitting parameters for the longitudinal conductivity and the Hall conductivity of HfTe$_5$ (sample 3) as shown in Figs. 3(c) and (d) in the main text.